\documentclass[journal,final]{IEEEtran}
\usepackage[noadjust]{cite}

\usepackage{graphicx}
\usepackage{subfig}
\usepackage{algpseudocode}
\usepackage{algorithm}
\usepackage[cmex10]{amsmath}

\newcounter{inlineenum}
\renewcommand{\theinlineenum}{\alph{inlineenum}}
\newenvironment{inlineenum}
  {\unskip\ignorespaces\setcounter{inlineenum}{0}%
   \renewcommand{\item}{\refstepcounter{inlineenum}{\textit{\theinlineenum})~}}}
  {\ignorespacesafterend}
  
\newcommand{\var}{\texttt}
\newcommand{\assign}{\leftarrow}
\newcommand{\multilinestate}[1]{%
  \parbox[t]{\linewidth}{\raggedright\hangindent=\algorithmicindent\hangafter=1
    \strut#1\strut}}

\usepackage[symbol]{footmisc}

\newcommand\Tstrut{\rule{0pt}{2.6ex}}       % "top" strut
\newcommand\Bstrut{\rule[-0.9ex]{0pt}{0pt}} % "bottom" strut
\newcommand{\TBstrut}{\Tstrut\Bstrut} % top&bottom struts

\DeclareRobustCommand*{\IEEEauthorrefmark}[1]{%
  \raisebox{0pt}[0pt][0pt]{\textsuperscript{\footnotesize\ensuremath{#1}}}}
  
\begin{document}

\title{Reconfigurable Integrated Optical Interferometer Network-based Physically Unclonable Function}

% The author marks are weirdly numbered because I am using symbolic footnotes for clarity; Need to pick marks that were not used.
\author{

\IEEEauthorblockN{A.~Matthew~Smith\thanks{A.M.S. and H.S.J. shared equal contribution to this work.}\IEEEauthorrefmark{1} and H~S.~Jacinto\IEEEauthorrefmark{1}$^,$\IEEEauthorrefmark{2},~\IEEEmembership{Member,~IEEE}}

\IEEEauthorblockA
{	\IEEEauthorrefmark{1} Air Force Research Lab, Rome, NY 13440 \\
	\IEEEauthorrefmark{2} Boise State University, Boise, ID 83725}
}




\maketitle

\begin{abstract}

In this article we describe the characteristics of a large integrated linear optical device containing Mach-Zehnder interferometers and describe its potential use as a physically unclonable function. We propose that any tunable interferometric device of practical scale will be intrinsically unclonable and will possess an inherent randomness that can be useful for many practical applications. The device under test has the additional use-case as a general-purpose photonic manipulation tool, with various applications based on the experimental results of our prototype. Once our tunable interferometric device is set to work as a physically unclonable function, we find that there are approximately $\mathbf{6.85\times10^{35}}$ challenge-response pairs, where each challenge can be quickly reconfigured by tuning the interferometer array for subsequent challenges.

\end{abstract}


% \fi
\IEEEpeerreviewmaketitle

\section{Introduction} \label{sec:intro} % sec I

\IEEEPARstart{P}{hysically} unclonable functions (PUFs) or physical one-way functions (POWFs) have been suggested as a method to securely authenticate a networked device or remote user. Current state-of-the-art means of authentication begins with the usage of a classical secret key, or token, stored within a read-only memory (ROM). PUFs are of particular interest since they often form the basis of hardware primitives necessary to replace shared secret keys with a non-reproducible physical object or device.

Classical CMOS-based PUFs are physical primitives that utilize process fabrication variance to create unique POWFs. Unlike non-volatile memory, where information can be stored and read digitally, information in CMOS-based PUFs is directly extracted from inherent lithographic variation, making static PUFs impossible to be duplicated; even within the original manufacturing process \cite{herder14}. Other common forms of CMOS PUFs include arbiter PUFs \cite{Xu2015} that utilize delays to measure differences in transmission times of two competing pathways in order to generate a digital response, butterfly PUFs \cite{kumar2008,Guajardo2007} that examine output from a set of cross-coupled latches, and random-access memory (RAM) PUFs \cite{Holcomb2009} that are based on randomly distributed mismatches between two transistors where the repeatable start-up conditions of cells are treated as digital responses. 

The operating scheme for all types of PUFs remains essentially identical: Given a set of specific inputs, referred to as the challenge, a PUF will generate a unique output response. These inputs and corresponding outputs are known as the challenge-response pairs (CRPs). The manufacturer or user of the PUF enrolls the device's unique information by generating and recording all of the viable CRPs. The user can then verify the identity of the integrated (or remote) PUF, at a later time, by challenging the suspect device and comparing the response to the expected response. In this work, our challenge will be a set of randomly selected voltages applied to the device, $\bar{C}$. The response will be the normalized distribution of laser intensity in the output modes that results from those voltage settings, $\bar{R}$. The details of this scheme will be described in Section~\ref{sec:the_qpp}.

PUFs based on optical measurements have been proposed with differing operating mechanisms, where either the scattering of laser-light from bulk inhomogeneous media \cite{pappu2002}, multi-mode fiber \cite{Mesaritakis2018}, or non-linear interaction in specialized integrated devices \cite{grubel2017} are observed. One of the main reasons that electronic PUFs are commonly implemented into field programmable gate arrays (FPGAs) and other protected IPs is due to the electronic PUFs' ease of integration into the many existing CMOS-process devices, alongside their low size, weight, and power requirements. 

Optical PUFs often require non-trivial bulk optics and ancillary support such as micron-accurate positioning stages \cite{pappu2002} or bulk disordered materials \cite{Mesaritakis2018}. A more compact solution was conceived by Grubel et al. \cite{grubel2017} through the utilization of photonic integrated circuits (PICs), however, these PIC PUFs require a set of completely custom-designed devices for the sole purpose of use as a PUF. Here we propose that any large- and well-connected-enough array of linear interferometric devices can be used for both its designed purpose \textit{and} as an optical PUF. The ongoing development of large-scale PICs, and particularly large interferometric devices \cite{obrien2018,obrien2018sci,obrien2017,englund17,harris14}, along with the wide range of applications from general information processing \cite{englund17}, quantum key distribution \cite{obrien2017}, quantum optics \cite{obrien2018}, and even the development of deep-learning and optical neural networks \cite{shen17, carolan2017apparatus}, suggest that such linear PICs will become ubiquitous components in the future.
Analogous to the development of RAM-based PUFS, our circuit was not specifically designed to act as a PUF. We demonstrate that the large interferometric circuits now being developed by the authors and other groups \cite{obrien2018,obrien2018sci,obrien2017,englund17,harris14,shen17,carolan2017apparatus,harris18} have an additional application as a PUF.

In this work we describe a linear optical interferometric PIC which acts as a PUF. We demonstrate how a small sub-circuit operates as a weak PUF, but has the ability to further meet the criteria of a strong PUF. We show how the scaling of an integrated optical circuit intrinsically carries enough randomness from multi-input interference via adjustments of Mach-Zehnder interferometers (MZIs) to act as a practical PUF. 

The major advantage of using such a PUF is compatibility with existing CMOS fabrication for easy adoption with existing technologies. The PIC PUF's tight integration with other PIC devices, such as those used in quantum communication, allows for improved security over add-on alternative PUF IPs. Our PUF's performance, and large number of challenges, discussed in Section~\ref{sec:results}, give it excellent characteristics as a strong PUF. Finally, the ability to use \textit{any} large circuit of interferometers as a PUF, combined with the growing size and number of such circuits, suggest that our device could become an ubiquitous component in the near future.

\section{The Quantum Photonics Processor} \label{sec:the_qpp} % sec II
\begin{figure}[tbp!]
  \centerline{\includegraphics[width=.45\textwidth]{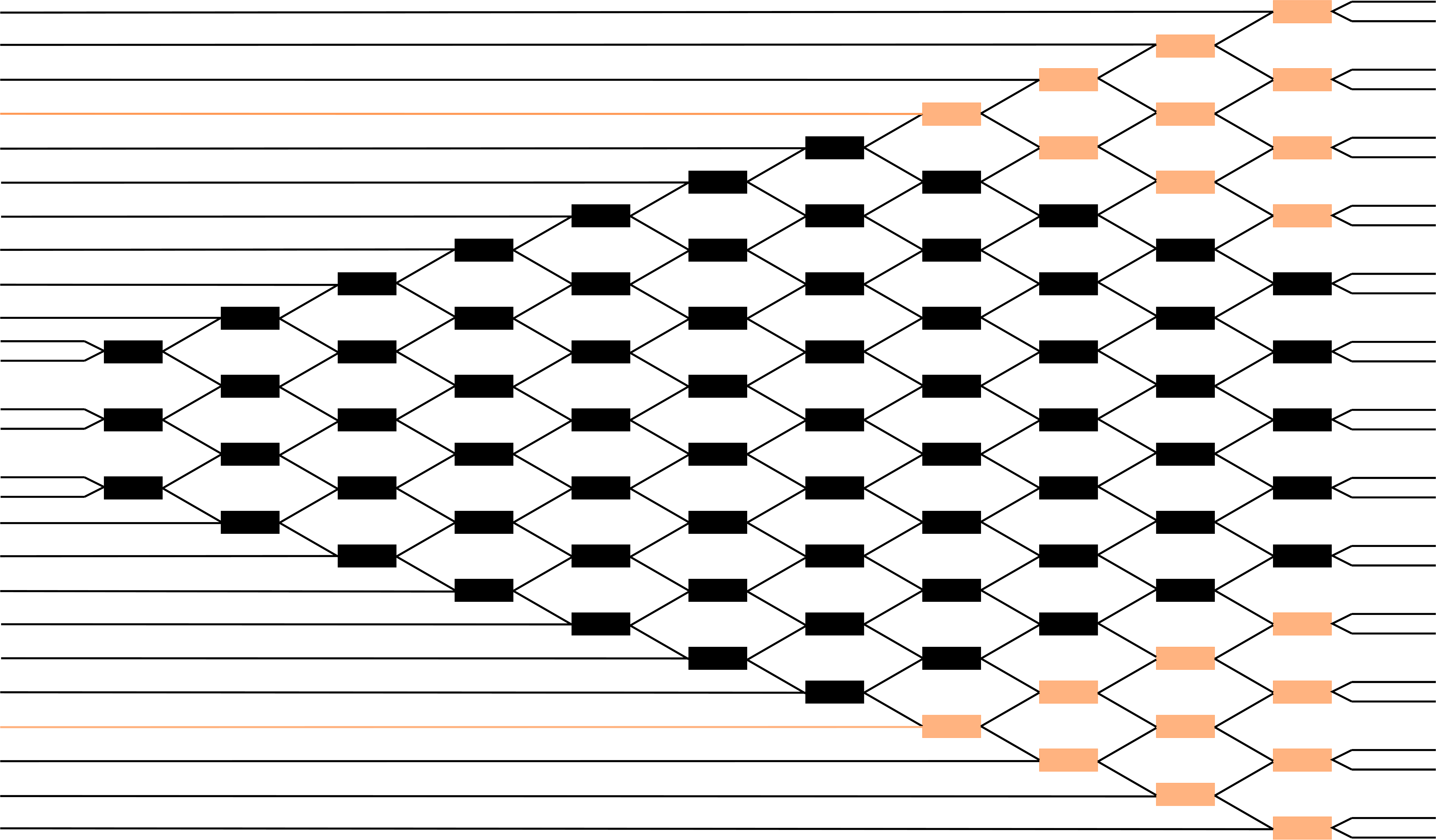}}
\caption{{\bf Structure of Quantum Photonics Processor}, where each rectangular box represents a single 2x2 MZI. The figure shows two logically separated devices (orange), consisting of 10 MZIs each, and the pumping scheme within the larger QPP. The largest PUF on the QPP is pumped down the center channel and consists of 66 MZIs.}
\label{diag1}
\end{figure}

In this research we have employed the Quantum Photonics Processor (QPP)\footnote{Also known as the Programable Nanophotonic Processor (PNP).} developed in collaboration between the Air Force Research Lab (AFRL) and the photonics research group, under the direction of Dirk Englund at MIT \cite{englund17,harris18}, as our prototype linear optical interferometric PUF. The optical PUF device is designed as a silicon-on-insulator (SOI) integrated optical chip fabricated in a CMOS foundry. The PIC device consists of 88 2x2 MZIs connected in a triangular nearest-neighbor configuration, as shown in Figure~\ref{diag1}.

\begin{figure*}[tbp!] 
  \includegraphics[keepaspectratio,width=\textwidth]{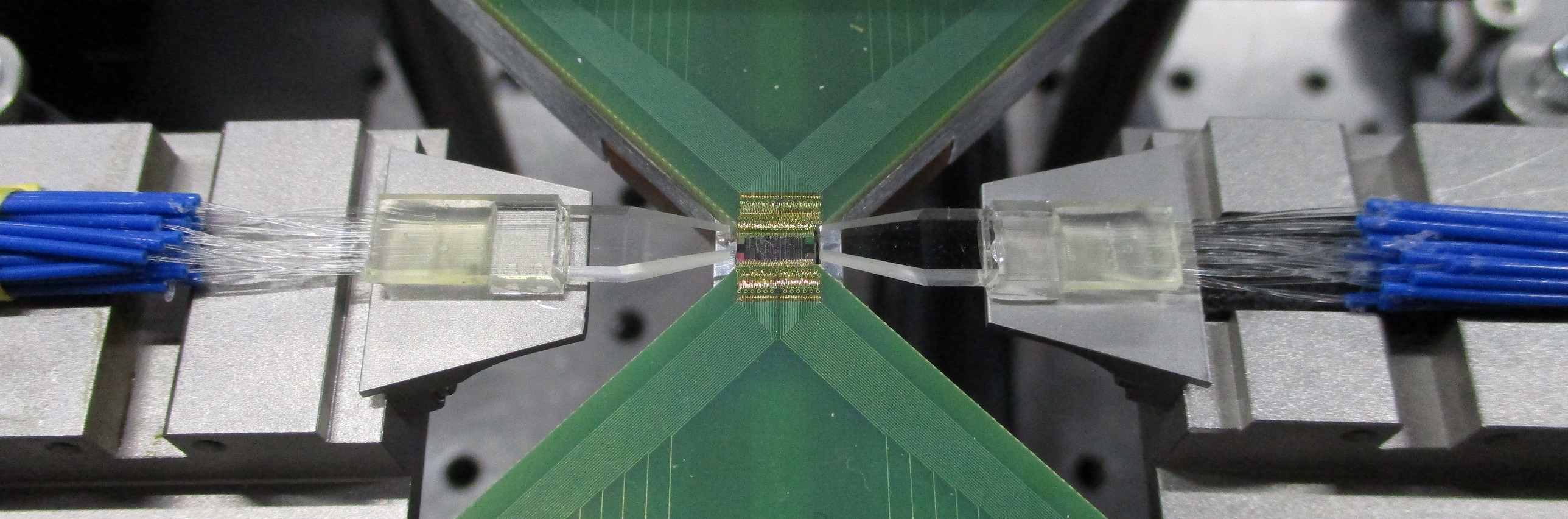}
  \caption{\textbf{Quantum Photonics Processor} shown with two 26-mode fiber arrays edge-coupled to the QPP. The fiber array on the left is the input of laser-light into the QPP, connected to a printed circuit board with connections to control the integrated resistive heating elements. The fiber array on the right leads to an array of PIN photodiodes to measure intensities on the outputs.}
  \label{fig:qpp_image}
\end{figure*}

The devices are standard Silicon on Insulator (SOI), a thermo-optic optic material, and as such each MZI can be independently, thermally, tuned by an integrated resistive heating element. The 2x2 MZI is capable of applying an idealized $2\times2$ unitary transformation shown in Equation~\ref{eq:mzi_xfrfn}, expanded upon in Section~\ref{sec:puf_metrics}. 
\begin{equation} \label{eq:mzi_xfrfn}
	U_{MZI}(\theta, \phi) = \dfrac{1}{2}
	\begin{pmatrix}
		e^{j\phi} & 0\\
		0 & 1
	\end{pmatrix}
	\begin{pmatrix}
		1 & j \\
		j & 1
	\end{pmatrix}
	\begin{pmatrix}
		e^{j\theta} & 0\\
		0 & 1
	\end{pmatrix}
	\begin{pmatrix}
		1 & j\\
		j & 1
	\end{pmatrix}
\end{equation}
Each MZI consists of two integrated phase shifters: One phase shifter between the beam splitters and a second phase shifter on a single output-leg. The unitary transformation in Equation~\ref{eq:mzi_xfrfn} includes two variables, $\theta$ and $\phi$, which map to the internal phase setting and output phase offset, respectively, for each MZI. In this work we only employed the internal phase shifters, driven by a computer-controlled voltage driver \cite{harris14}, and thus have two PUF devices consisting of 10 MZIs, pumped by a Keysight laser (model 81606A) through a single waveguide, respectively. The fiber-arrays and edge-coupling of the QPP is shown in Figure~\ref{fig:qpp_image}. Each PUF device has 8 output ports, each connected via SMF-28 to a single standard PIN photodiode (Precision Micro Optics model DPRM-412). The subset devices used within the QPP are triangular-shaped with a light-cone-like dispersion region, visualized by the orange regions in Figure~\ref{diag1}.

The first source of randomness for this device is the $\approx15.43\%$ variation between resistive heating elements, as measured, due to fabrication variances. It should be noted that the continual operation of the resistive heating elements will lead to electrode annealing, thus a change in the output of the PUF could be observed over time. A far more significant effect on randomness are the two directional couplers within each MZI. The couplers are designed to be a nominal 50:50 splitting ratio but fabrication defects stemming from variation in the etching process, sidewall roughness, and variation in minute distances between waveguides leads to unpredictable splitting ratios near $50\%$. An additional source of unpredictability leading to potential for randomness in the device comes from a minor design flaw: Since many of the MZIs share ground leads, positive feedback ground-loops are formed when a single MZI's voltage is set and the cascading MZI's resistive elements return a complex set of voltages, set by nearest-neighbor association. The effect of ground-loop feedback is approximately $-45$ dB as measured by M. Prabhu \cite{prabhu2018towards}. It can be expected that the positive feedback ground-loop voltage errors may be a minor factor in the device's overall behavior. To minimize unwanted global thermo-optic effects, the device was held at a steady temperature, slightly above ambient, throughout testing. The effects of operating at differing temperatures have not yet been explored, however, the device can operate at a wide range of steady temperatures, with each having an impact on output characteristics, without damage. The differing global temperature effects on the PUF are further discussed in Section~\ref{sec:results}. 

The QPP is large enough to act as two distinct devices with identical structure. Two devices were programmed to be used for comparison by taking the QPP and pumping laser-light into two space-like separated regions such that the light from one device will not reach the other device, either directly or through reflections other than those coupled into the slab-mode. In addition, the two devices are electrically separated so that no positive ground-loop feedback effects exist between the devices. 

\section{PUF Metrics} \label{sec:puf_metrics} % sec III

We use the definitions of a weak and strong PUF given by C. Herder et al. \cite{herder14}. A weak PUF is described as:
\begin{inlineenum}
	\item Having a number of CRPs linearly related to the number of components,
	\item being robust against environmental effects, i.e. having stable CRPs,
	\item having unpredictable responses to any stimulus and,
	\item being extremely impractical to reproduce. A strong PUF is characterized by all of the previous statements regarding weak PUFs with the addition of:
	\item Having enough CRPs such that the number is exponential in the number of challenge bits and
	\item that the readout will reveal only the response $\bar{R}=f(\bar{C})$ and nothing else.
\end{inlineenum}

One metric chosen to test the difference between CRPs is the Euclidean distance, $\ell^2$-norm, of the $N$ outputs. To measure the Euclidean distance the analog response of each detector is divided into even-sized subsets; each of which is larger than the estimated noise of the \textit{system}. For our test we chose a subset of size $0.5\%$ of the total power detected across the $m$ outputs, scaled by a cross-normalization factor between CRPs.

To decrease and/or correct error within the testing of our PUF the size of voltage subset utilized in computation was increased to $0.1\%$. The increase in subset size serves to decrease the chances that any noise present on a particular channel straddles the bounds between two values. The increase in subset size also applies a reduction in resolution for the $\ell^2$ distance. An alternative option to decrease and/or correct error within the testing of our PUF is to increase the collection time, thus increasing the amount of sample-averaging that forms a single CRP. The drawback to relying on increased collection times are the latency requirements, which may hamper any fast electronics requiring the output of the PUF, and the possibly of allowing an adversary additional time to perform side channel attacks. 

The second set of metrics used to quantify the results and operation of the PUF are the inter- and intra-device Hamming distances ($HD_{inter},HD_{intra}$) along with the inter- and intra-device Euclidean norms ($\ell^2_{inter},\ell^2_{intra}$). To analyze the results we modified the standard Hamming distances between a response $\bar{R}_i$ from challenge $\bar{C}_i$, and response $\bar{R}_j$ from challenge $\bar{C}_j$, to reduce the effects of noise. The loose Hamming distance ($LHD$) can be calculated between two noisy response vectors, $\bar{R}_i$ and $\bar{R}_j$, for all of the corresponding laser-light intensity measurements on the output modes $k$ as: 
\begin{eqnarray}
\label{modham}
LHD=\sum_{k}{f(\bar{R}_i,\bar{R}_j)_k}=\begin{cases}
			0,  \forall k~\text{if}~|R_{i,k}-R_{j,k}|< L\\
			1,  \forall k~\text{if}~|R_{i,k}-R_{j,k}|\geq L\\
			\end{cases}
\end{eqnarray}
Where $L\in \mathbf N$ defines the degree of looseness, where $L=1$ is the normal Hamming distance. For the case of small PUFs, $L=2$ is sufficient. The definition for $LHD$ is used to compensate for the experimental noise and rounding errors as discussed below. In addition to the $LHD$, the standard $\ell^2$-norm is is used following the standard definition given by:
\begin{equation} \label{l2norm}
	\|\bar{x}\|_2=\sqrt{\sum_i{x_i^2}}.
\end{equation}

The major difference between these two metrics, for non-binary data, is that the Hamming distance represents the number of measurements which are different, while the Euclidean norm gives a metric of the significance of differentiation. Interestingly, we can expand upon the Hamming distances and determine the \textit{uniqueness} of our device as described by R. Maes et al. \cite{maes2010physically}. Uniqueness is a calculated estimate for the amount of entropy available from a PUF and can be applied to a similar population of PUFs with an identical architecture. The uniqueness, $\mathcal{U}$, can be calculated for some challenge $\bar{C}_i$ as:
\begin{equation} \label{eq:pufunique}
\mathcal{U}|_{C_i} = \left(\dfrac{2}{n(n-1)} \sum\limits_{i=1}^{n-1} \sum\limits_{j=i+1}^{n} \dfrac{LHD(\bar{R}_i, \bar{R}_j)}{m}\right) \times 100\%\,,
\end{equation}
Analogous to Equation.~\ref{modham}, $L=1$ gives the standard definition of uniqueness. Here, $n$ is the number of PUFs in a population, and $m$ represents the number of bits in the response from the PUF. An optimal uniqueness value for binary PUFs would be $50\%$, as this implies uncorrelated responses. Since our PUF is continuous via electronic control, we need to modify our interpretation of Equation~\ref{eq:pufunique}. Given that $LHD=0$, i.e. a complete collision, doesn't contribute to $\mathcal{U}|_{C_i}$ and a partial collision contributes only the fraction that didn't collide, $\mathcal{U}|_{C_i}$ is counting non-colliding responses. Regardless of the looseness this is equivalent to a target uniqueness between devices of $100\%$.

\section{Results} \label{sec:results} % sec IV
For this work we created several sets of data. First, using the two small sections from Figure~\ref{diag1}, we created 100,000 random CRPs mirrored on each device, and an additional single CRP was repeated 5,000 times on each device. Secondly, we tested the largest PUF that would fit on the device, again with random CRPs, and a repeated CRP. All of the CRPs were randomly selected in each variable from a uniform distribution over the $v2\pi$ voltage range required for a complete switching response of a typical MZI, detailed by the sinusoidal response from Equation~\ref{eq:mzi_xfrfn}, easily modified into a sine/cosine format shown in Equation~\ref{eq:mzi_xfrfn_sincos}.

\begin{align} \label{eq:mzi_xfrfn_sincos}
	U_{MZI}(\theta, \phi) =&\dfrac{1}{2}
	\begin{pmatrix}
		e^{j\phi}(e^{j\theta}-1) & je^{j\phi}(e^{j\theta}+1)\vspace{3pt}\\
		j(e^{j\theta}+1) & -(e^{j\theta}-1)
	\end{pmatrix}
	\nonumber\\= &je^{\frac{j\theta}{2}}
	\begin{pmatrix}
		e^{j\phi}\text{sin}\big(\frac{\theta}{2}\big) & e^{j\phi}\text{cos}\big(\frac{\theta}{2}\big)\vspace{3pt}\\
		\text{cos}\big(\frac{\theta}{2}\big) & -\text{sin}\big(\frac{\theta}{2}\big)
	\end{pmatrix}
\end{align}

\subsection{A Small PUF}\label{sec:smallresults} % sec IV.A.
To complete the analysis of responses of the small PUFs depicted in Figure~\ref{diag1}, eight output intensities were measured via a polled array of photodiodes for each device, with results below.

\begin{figure}[tbp!]
\centering
\subfloat[Device 1]{{\includegraphics[width=.25\textwidth]{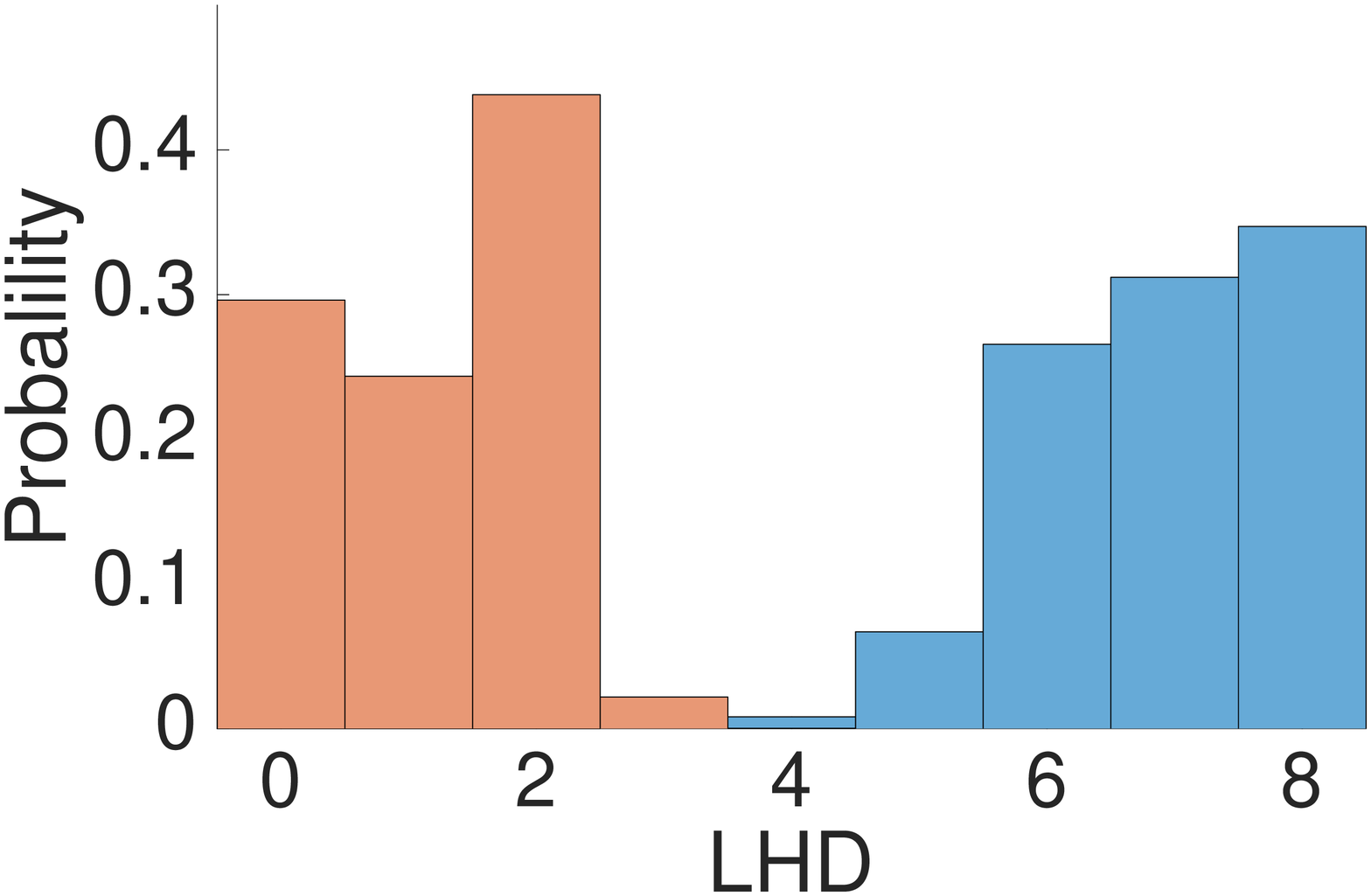}}}%
\subfloat[Device 2]{{\includegraphics[width=.25\textwidth]{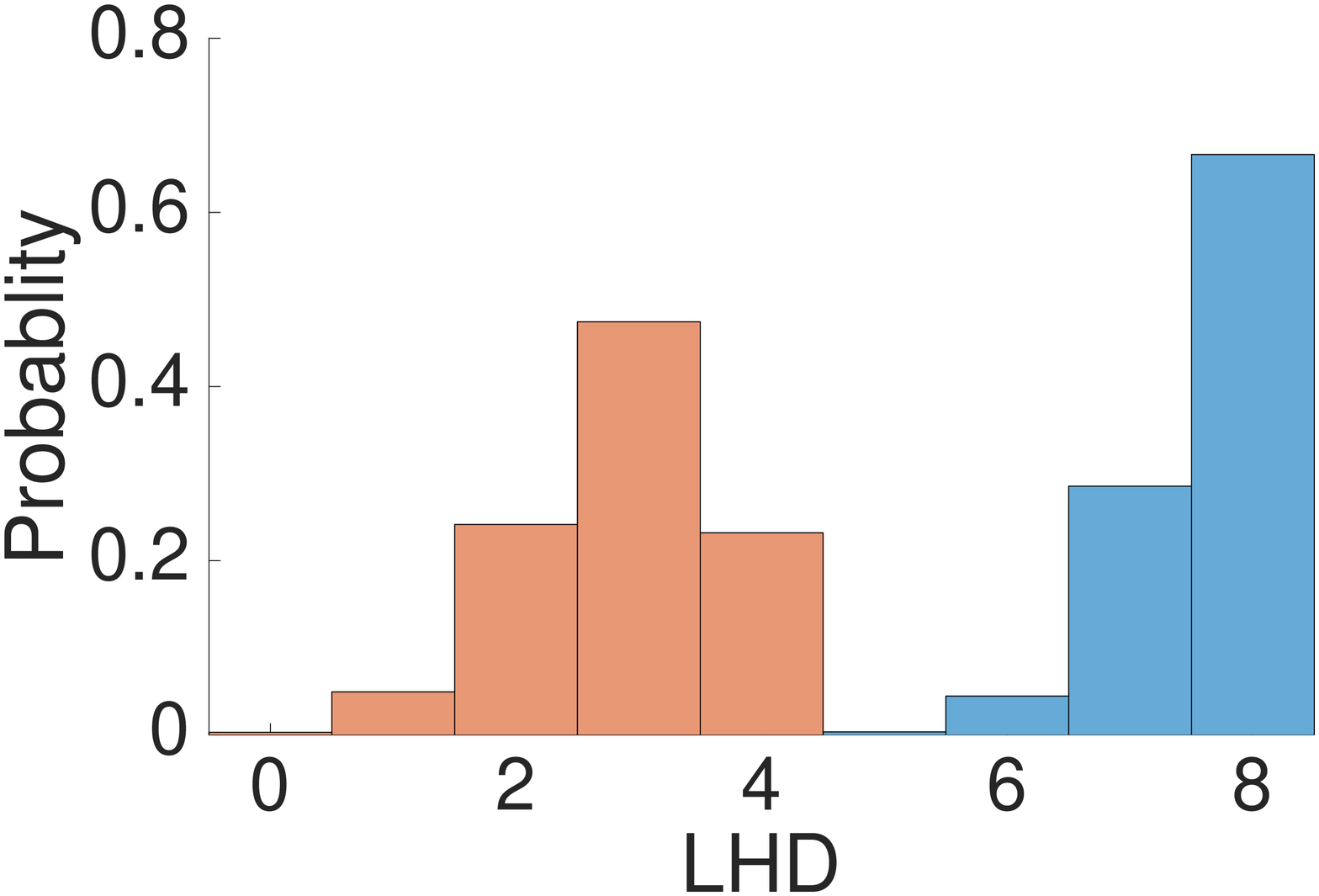}}}%
\caption{{\bf Distinguishability of $\mathbf{ LHD_{intra}}$}, for both devices. $LHD_{intra}$ between the same repeated challenge (orange) and between a typical challenge and random challenges (blue) on the same device. }
\label{intra}
\end{figure}

Figure~\ref{intra} shows the repeatability (orange) of the same challenge applied 5,000 times to each device. The two devices show a relatively low $LHD \leq 4$. The difference between Figure~\ref{intra}a and Figure~\ref{intra}b is accounted for by the differences in noise level, with a higher total noise on the second device\footnote{This difference is likely caused by photodetector variation due to differing production batches with a result of approximately 1.5 times the noise shown on the datasheet for the PIN photodiodes previously mentioned.}. The second dataset in both figures (blue) shows the difference between a typical CRP and the 100,000 randomly selected CRPs. The two devices show strong repeatability through the $LHD$ metric, by staying within a narrow variable range. The two devices additionally show strong metrics for distinguishability. The differences between a single CRP to a differing CRP set is easily identified. Ideally, $LHD_{intra}=0$ should be true for a fixed challenge, and $LHD_{intra}=8$ for differing challenges. The $\ell^2$-norm is necessary to provide an additional measure of the significance of differences before validating the identity of a PUF.

\begin{figure}[tbp!]
  \centerline{\includegraphics[width=.32\textwidth]{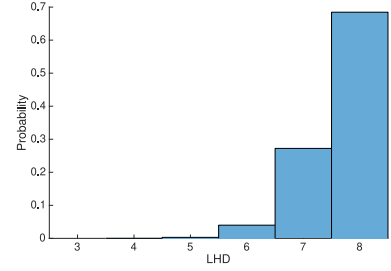}}
\caption{{\bf $\mathbf{LHD_{inter}}$}, Distances between 100,000 randomly chosen challenges applied to two near-identical devices, with the comparison between the two resulting responses, showing no full collisions.}
\label{inter}
\end{figure}

For applications of this PUF in authentication roles, the key importance is the inter-chip response to the same challenge. Figure~\ref{inter} depicts the $LHD_{inter}$ metric between 100,000 randomly chosen CRPs as they apply to both devices. As discussed further, the number of challenges is too large to test all possible settings. For 100,000 challenges mirrored between the two devices, we analyze Equation~\ref{eq:pufunique} to find a total uniqueness of $85.28\%$ at $L=2$. $LHD_{inter}$ is strongly centered around $LHD_{inter}=8$, approximately $70\%$ the response vectors have no corresponding measurement values in common ($R_{i,k}\neq R_{j,k}, \forall k$), and less than $10\%$ have more than two out of eight corresponding measurements in common. We found no complete collisions\footnote{In this context we take a collision to be complete where all output values are identical for two different given challenges, i.e. $\bar{R}_i$ and $\bar{R}_j$  where $R_{i,k} = R_{j,k}, \forall k$, or partial where two challenges result in some similar outputs; these are not fully distinguishable since a distinguisher can exist where a value can be differentiated from a random oracle \cite{bellare1993random}.}, which would be required for a false positive identification. This is highly encouraging as our devices are as identical as physically possible due to simultaneous fabrication (identical material, design, processing, environment, etc.) and are effectively `clones' of one-another. Any attempt to physically recreate a device to be utilized as a PUF will inherently possess additional random variance and produce a clearly differentiated $LHD_{inter}$ versus the original device.

\begin{figure}[tbp!]
  \centerline{\includegraphics[width=.5\textwidth]{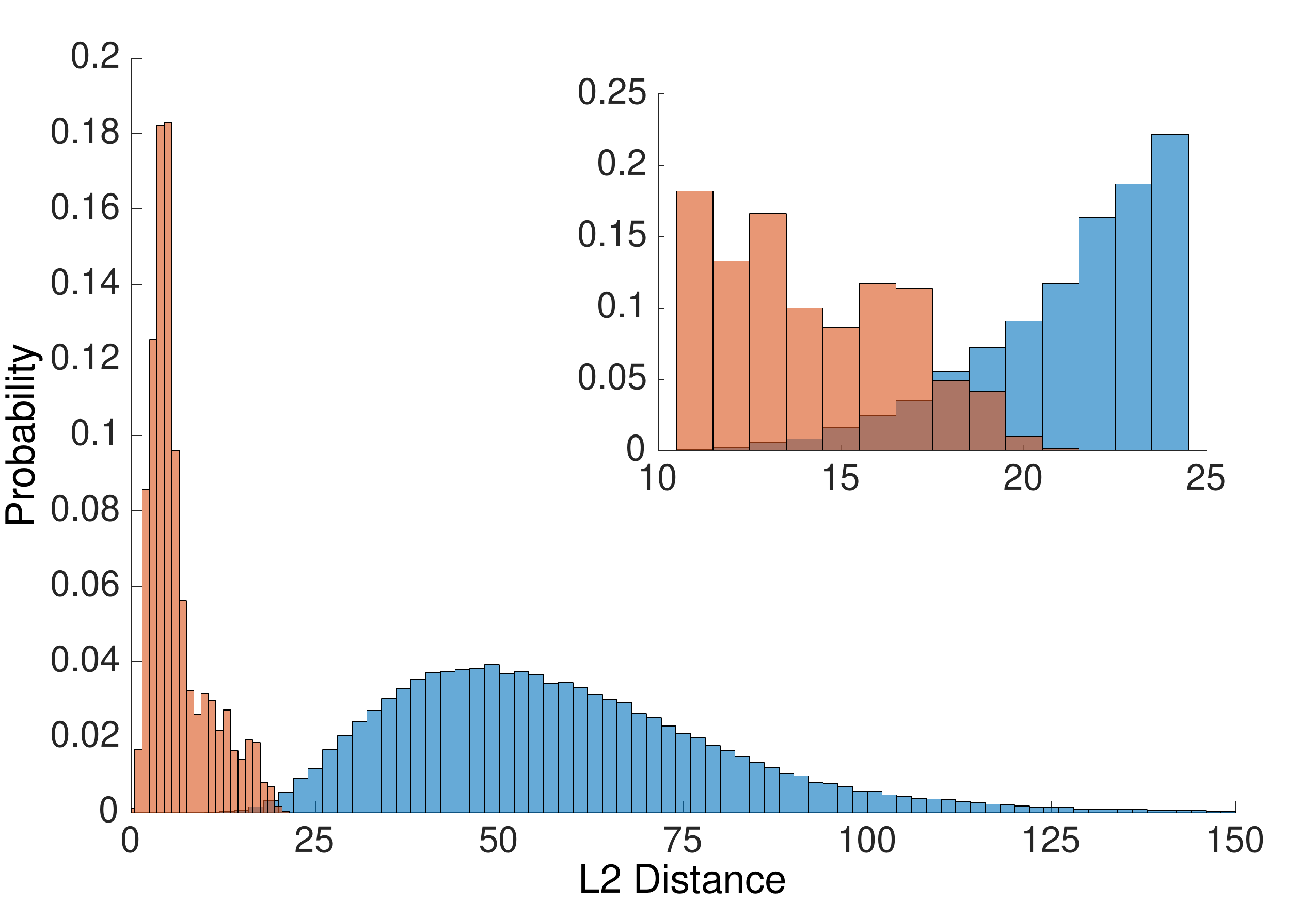}}
\caption{{\bf Euclidean distances}, showing the distance between the response to identical voltage settings on both devices ($\ell^2_{inter}$, blue) and the response of one device to the same repeated challenge ($\ell^2_{intra}$, orange, typical).  The inset shows the region of overlap.}
\label{l2dist}
\end{figure}

The commonality of the measurements are shown in Figure~\ref{l2dist}, where the blue data shows the $\ell^2_{inter}$ distance between the two devices, for each challenge. The smallest $\ell^2_{inter}$ distance found was 11, with a mean of 58, median of 55, and standard deviation of 23. The orange data shows the $\ell^2_{intra}$ distance between a typical response and all other responses to the same challenge on a single device. The data shown is typical with limited overlap between histograms. Some CRPs appear to have more noise than others, and multiple datasets have shown no overlap at all between histograms, the least distinguishable of which is shown as an example in Figure~\ref{l2dist}; the $\ell^2_{intra}$ data shown here has a mean of 6, median of 5, and a standard deviation of 4.

A major source of errors during measurement, unfortunately, is the instability of input- and output-coupling, leading to variability in total intensity over time. There are two significant components to the coupling errors; a high-frequency component and a slow drift caused by undamped environmental noise and sagging positioning stages holding the edge-coupled fiber arrays, respectively. The signal to noise ratio (SNR) within our system is approximately $16$ dB. The laser and detector SNR is estimated to be approximately $50$ dB, based on the dark-count of the photodetectors. The losses on the integrated chip are stable and do not vary with time. The PIN photodiodes used to measure the output signals have a constant background noise and gain, such that the effects of losing overall intensity represents a reduction in SNR. For our comparison, the loss of intensity can be counteracted by normalizing the total intensity and the set-sizes over each CRP to a fixed value prior to calculating the $\ell^2$-norm. As such, the result of comparison is the difference between relative intensities for each channel and not the total intensities. The main source of the intensity error is a result of physically edge-coupled fiber arrays, on manual positioning stages, rather than permanently affixed arrays. A packaged device with permanently affixed fiber arrays are a requirement for any practical system and will nearly eliminate this source of error.

\subsubsection{Range and Number of CRPs}\label{subsec:recurrence_count} %IV.A.1. 
The total significance of any individual setting within a CRP is not uniform across the device. Of the ten input variables in each small device, four can only affect two of the output measurements each in the current design. Physically, this can be observed as the last column of 4 MZIs or the base of the pyramid in Figure~\ref{diag1} as opposed to the input or `tip' of the pyramid that affects all output channels. A concern with this architecture is the possibility that a nearest-neighbor challenge will produce a semi-predictable response and there are many fewer distinguishable responses than there are challenges.

Therefore, an important question for any proposed PUF is how many uniquely identifiable CRPs are available and how the number of CRPs scales with the size of PUF. Weak PUFs can have as few as one CRP, although this is a severe limitation on the number of applications \cite{herder14}. Given the input amplitudes and detector noise on the output modes, we are able to clearly distinguish a voltage change on a single MZI of $\approx7mV$. As each MZI in the system has a $v2\pi\approx$7V range, we define 10 bits of resolution in voltage as $v2\pi/1024$ for each MZI. Na\"ively we could estimate that since the PUF has 10 bits of resolution, on ten independent MZIs, as shown in Figure~\ref{diag1}, there are $2^{10^{10}}\approx 1.27\times10^{30}$ possible challenges on each device. However, this ignores a large number of partial collisions in the output space of the PUF and also ignores the structure of the device. When taking the largest set of MZIs in a light-cone pattern, we see that, at most, 66 MZIs exist in our architecture. Assuming a 10-bit resolution, we could again theorize a maximal upper bound of challenges for some set of 66 MZIs of $2^{10^{66}}\approx 4.78\times10^{198}$. 

However, this calculation is out of scope: Determining the number of challenges within our architecture that have \textit{differentiable} responses is a more difficult and productive task. For our architecture, with a light-cone diffusion, we can better approximate a maximal upper bound by following the Catalan numbers \cite{koshy2008catalan}, $C_n$, from combinatorics and count the number of \textit{distinguishable} settings by analyzing the MZI structure as a fully-rooted binary tree with $n+1$ leaves. We can use a rooted binary tree since we pump the PUF from a single input and can calculate an upper bound given by:

\begin{equation}
\label{catalan}
C_n=\frac{(2n)!}{(n+1)!n!}.
\end{equation}

Following the Catalan numbers, we calculate that for 10-bit resolution there are approximately $5.77\times 10^{39}$ combinations for an array of 66 MZIs, not necessarily our specific configuration of MZIs. This maximal upper limit is unfortunately still too large due to the number of configurations of 66 MZIs that are not possible within our architecture. The limit of the architecture where pure Catalan numbering cannot apply is due to the limited number of columns in our device. The limit in number of columns means that the Catalan numbering scheme will count combinations in a light-cone pattern that are impossible: Think of an arrangement where the first 11 MZIs are set to pass light linearly down in a straight line with an additional 55 MZIs hanging off of the end of our architecture.

\begin{algorithm}[htbp]
	\caption{Algorithm to calculate Catalan recurrence for planar trees $t_i(n, h)$.}
	\label{alg:catalanrecurr}
	\begin{algorithmic}
		\Require $n \geq 0$ and $h > 0$
		\State \multilinestate{$\var{recurs} \assign{}$ values of \var{n} and \var{h} into a 2 dimensional array at index [\var{n}, \var{h}] with the number of combinations.}
		\State \multilinestate{$\var{sum} \assign{}$ values of operations and calculates the sum.}	
		\Procedure{Recurrence}{[$n$, $h$]}
			\State $i$, $j$, Local Variables
			\State $n^\prime$, $h^\prime$, Shadow Copy of $n$ and $h$
			\State \var{recurs}[0, 0] = 1
			\If {$h < n$ or $h>2^n-1$}
				\State \var{recurs}[$n^\prime$, $h^\prime$] = 0
			\ElsIf {$1 \leq n \leq h \leq 2^n-1$}
				\For {$i$, $j$}
					\State \var{recurs}[$n^\prime$, $h^\prime$] = \var{recurs}[$n$, $h$] 
					\State = \var{sum}[\var{recurs}[$n-1$, $h-1-i$]
					\State \quad$\times$(2$\times$\var{sum}[\var{recurs}[$j$, $i$], ($j$, 0, $n-2$)]
					\State \quad+ \var{recurs}[$n-1$, $i$]), ($i$, 0, $h-1$)]
				\EndFor
			\EndIf
		\EndProcedure
	\end{algorithmic}
\end{algorithm}

To overcome the configuration limit set by the standard Catalan numbers, we must utilize a lesser-known combinatorics counting method for binary trees as described by F. Qi et al. \cite{qi2017integral}, the method of counting by integral representation of the Catalan numbers. The method of integral counting can be directly applied to the planar tree variation of counting, similar to the work by P. Flajolet et al. in \cite{flajolet1982average}.

If, for a forest composed of a set of trees, $\mathcal{F} = \{t_0, t_1, \ldots, t_k \}$, we look at a single tree, $t_i(n, h)$, this tree can represent any binary tree with or without a shared child of height $h$ with $n$ nodes. Simply, $\sum_h\,t_i(n,h)=C_n$, for the $n$-th Catalan number. By analysis, the Catalan recurrence for a planar tree gives the recurrence formula for $t_i(n, h)$\footnote{This formula requires the following definitions $t_i(0,0)=1$ and $\forall \, t_i(0,-)=0$.}:
\begin{align}\label{eq:recurrence}
	t_i(n+1, h+1) 	&= 2 \sum\limits_{m=h+1}^{n} t_i(m,h) \sum\limits_{j=0}^{h-1} t_i(n-m, j)\nonumber\\ 
					&+ \quad \sum\limits_{m=h+1}^{n-h-1} t_i(m, h)\,t_i(n-m, h)\, .
\end{align}

The formula in Equation~\ref{eq:recurrence} utilizes the double summation to count the number of combinations to build a binary tree on $n+1$ vertices whose left sub-tree has a height $h_0$, and whose right sub-tree has height $h < h_0$. Doubling this value by a factor of 2 adds all trees whose right sub-tree have height $h^\prime_0$, and whose left sub-trees have height $h^\prime < h^\prime_0$. The final term of Equation~\ref{eq:recurrence} serves to count the planar trees on $n+1$ vertices whose left and right sub-trees are of height $h$.

To adequately enumerate the number of distinguishably different CRPs, Algorithm~\ref{alg:catalanrecurr} was used to implement Equation~\ref{eq:recurrence} with parameters $0 \leq n \leq 11$ and $0 \leq h \leq 66$. After running Algorithm~\ref{alg:catalanrecurr}, the total number of distinct CRPs that are possible with 10-bit resolution are calculated to be $\approx 6.85\times 10^{35}$. The full trees of MZIs configured into different columns with 10-bit resolution have the number of possible configurations shown in Table~\ref{tab:configs}. It should not be a surprise to see that as the number of columns increases, the possible configurations increases up to a point of maximal dispersion for sub-trees. The information regarding number of configurations due to architectural change will not be discussed further in this work, but may pose as an interesting topic of research.

\begin{table}[htbp!]
\caption{Variation in number of configurations versus number of columns for light-cone configurations within our device.}
\label{tab:configs}
\resizebox{\columnwidth}{!}{
\begin{tabular}{cc}
\hline
\multicolumn{1}{|c|}{\textbf{Number of (Columns, MZIs)}} & \multicolumn{1}{c|}{\textbf{Possible 10-bit Configurations}}\TBstrut\\\hline\hline
\multicolumn{1}{|c|}{(4, 10)}                         & \multicolumn{1}{c|}{$1.19 \times 10^{5}$}          \TBstrut\\ \hline
\multicolumn{1}{|c|}{(5, 15)}                         & \multicolumn{1}{c|}{$2.40 \times 10^{7}$}           \TBstrut\\ \hline
\multicolumn{1}{|c|}{(6, 21)}                         & \multicolumn{1}{c|}{$2.76 \times 10^{10}$}          \TBstrut\\ \hline
\multicolumn{1}{|c|}{(7, 28)}                         & \multicolumn{1}{c|}{$1.61 \times 10^{14}$}          \TBstrut\\ \hline
\multicolumn{1}{|c|}{(8, 36)}                         & \multicolumn{1}{c|}{$4.40 \times 10^{18}$}          \TBstrut\\ \hline
\multicolumn{1}{|c|}{(9, 45)}                         & \multicolumn{1}{c|}{$5.43 \times 10^{23}$}          \TBstrut\\ \hline
\multicolumn{1}{|c|}{(10, 55)}                        & \multicolumn{1}{c|}{$2.94 \times 10^{29}$}         \TBstrut \\ \hline
\multicolumn{1}{|c|}{(11, 66)}                        & \multicolumn{1}{c|}{$6.85 \times 10^{35}$}         \TBstrut \\ \hline                                                                    
\end{tabular}}
\end{table}

\subsection{A Large PUF} \label{subsec:largepuf} % sec IV.b.
The experiment above was repeated with the largest available PUFs on the current design of our chip. This was a `pyramid' consisting of 66 MZIs, with one input mode and 22 outputs. Here the `top' and `bottom' individual PUFs are so large as to significantly overlap, in fact they share 45 out of 66 MZIs. Despite this overlap, we find excellent distinguishability between the two PUFs. Not surprisingly the distinguishability is improved over the small PUFs. However, there is a limit to the improvement with size. As the total intensity is distributed over an increasingly larger number of outputs, the SNR will limit the maximum size with any given laser power and sources of loss.

\begin{figure}[tbp]
  \centerline{\includegraphics[width=.5\textwidth]{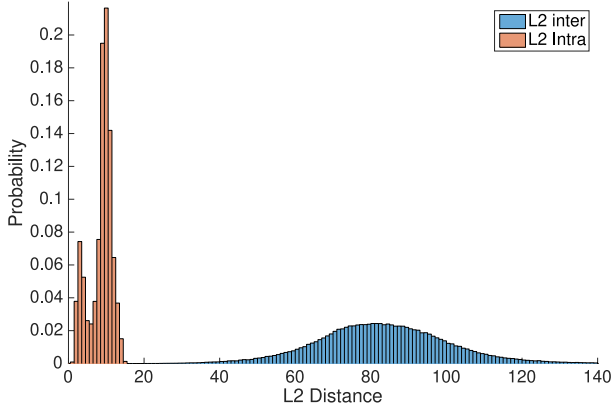}}
\caption{{\bf Euclidean distances of the large PUFs}, showing the distance between the response to identical voltage settings on the large devices ($\ell^2_{inter}$, blue) and the response of one device to the same repeated challenge ($\ell^2_{intra}$, orange, typical). The two peaks in the $\ell^2_{intra}$ are most likely the result of unstable coupling combined with our normalization method, creating two distinct noise levels during data collection. } 
\label{l2distbig}
\end{figure}

Figure~\ref{l2distbig} shows the calculated $\ell^2$-norms, similar to Figure~\ref{l2dist}. Note that the lack of an inset is the result of no overlap between the histograms. Increasing the width of the interferometer array is a route to increasing the total number of CRPs, as our current chip has three such 66 MZI subsets, giving a total number of CRPs on our device of $\approx 2.05\times 10^{36}$. Simply adding another row of just 11 MZIs appears to add to $\approx 6.85\times10^{35}$ distinct CRPs.

\begin{figure}[tbp]
  \centerline{\includegraphics[width=.5\textwidth]{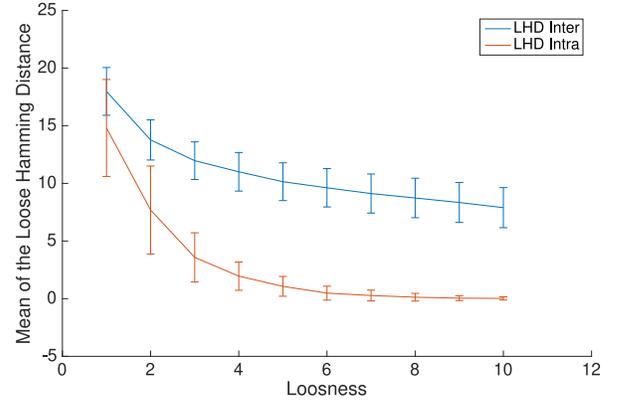}}
\caption{{\bf Mean of the LHD with looseness}, showing the difference between the mean of the $LHD$ for different and repeated challenges at various definitions of the looseness parameter $L$}
\label{loose}
\end{figure}

One question we asked is in regard to the looseness of the Hamming Distance, $LHD$, and what the ideal degree of looseness would be for the system. Recall that by `loose' we refer to a Hamming distance that overlooks small errors due to the noise and instability in our system. Naturally, the ideal parameter will vary based on the nature of the noise in the distributions being compared. We calculated the Hamming distance with varying degrees of looseness from the strict definition of Hamming distance ($L=1$) to $L=10$, and found the mean on resulting probability distribution; results shown in Figure~\ref{loose}. Note that the significant overlap of the first standard deviation error bars at $L=1$ results in the two PUFs looking all but identical under the strict Hamming distance definition. Under the $\ell^2$-norm and $LHD$ there is no measured overlap in Figure~\ref{l2distbig} or Figure~\ref{loose}, respectively.

The mean of the $LHD$ asymptotically approaches zero for for both test cases, however, the repeated challenge (orange) approaches significantly faster than the random comparisons (blue) in figure Figure \ref{loose}. The optimal parameter for the looseness was found to be between 5 and 6. At this point the two means are separated by 5.5 standard deviations of $LHD_{Intra}$, implying clear differentiability. This $LHD$-based approach is general, and as-such, this form of analysis can be used for any PUF with noisy output. 

\subsection{Attributes of an Optical CMOS-Compatible PUF} %sec IV.c.

The results shown in Section~\ref{subsec:recurrence_count} serve to highlight the optical interferometric PUF's ability to scale exponentially, thus meeting the first criterion for a strong PUF by C. Herder et al. \cite{herder14}. An additional facet of the design shown is the ability to have quick reconfigurability to assess additional CRPs. Since each of the MZIs are independently tunable, we can see the response of a tuned device and change parameters for subsequent CRPs. The ability to tune our device at-will enhances application and use-cases to not only the static processing of information, but the processing of streaming information. It is thus possible to process information streamed through the device or static information where a set of CRPs is dynamically changed depending on the information received.

Unfortunately, there are several negative attributes to using a system of interferometers as a PUF. As mentioned in Section~\ref{sec:puf_metrics}, the nearest-neighbor challenges may give predictable results on smaller PUFs. In addition, the interferometric system is highly structured and fixed, such that a sufficient number of CRPs being calculated could lead to the device being fully characterized. Indeed the QPP was designed with such characterization in mind as the original use case was for applications and experimental testing of quantum optical networks \cite{mower2015high}. We point out that the device was not originally intended to act as a PUF and we are merely exploiting its attributes. Since the reconfigurability of the QPP is available, it is possible to make one device clone the function of another device; for purposes as a PUF it is suggested to utilize this device in an uncalibrated mode. Custom designed interferometric circuits with more complicated interconnections, including variable feedback loops, would be more resistant to characterization and thus act as stronger PUFs.

A second negative attribute of the current prototype is that the operating temperature must be stable within $\pm 1^\circ$C. Allowing the temperature to vary may be a route to increasing the number of challenges and response pairs (CRPs). Whether or not two devices respond differently to temperature changes based on CRP normalization is an open research question. If temperature variation were not desired, packaging the device may lead to an easy method of stabilization. Alternatively, multiple sets of CRPs can be created for an array of temperatures prior to use. The variation with temperature observed is a direct result of using common SOI and CMOS fabrication. Silicon is a thermo-optic material and was chosen for its ease of integration into existing CMOS processes. However, the design for an interferometric optical PUF can be trivially transferred to an electro-optically controlled material such as Lithium Niobate to create a more stable standalone device, or application specific integrated circuit. It should be noted, however, that Lithium Niobate will still have a small thermo-optic effect. Conversely, each challenge $\bar{C}_i$ could double at as bias setting for the device. Variation in other parameters like global heating of the device, wavelength inputs, and variance in the number of pumped channels can allow each challenge to be utilized as an individual, separate, PUF. Here we have taken these parameters to be constants for simplicity but if allowed to vary, utilizing more parameters opens an enormous space of possibilities, and significantly increases the number of CRPs theoretically available.

Finally, with the software drivers used in these experiments we take approximately 3 seconds to completely set a challenge and measure a response of 1,000 physical measurements on the QPP. This has since been significantly improved with new driver optimization. The fundamental limit to the speed of the challenge and response is set by the maximum speed that the thermal switching can occur; estimated to be in the $\approx 100 \text{kHz}$ range \cite{harris14}. This may appear slow but we stress that the experimental setup was in no way designed to optimize the speed of measurements. The system currently runs on several standard Arduino-driven Teensy boards, for ease of development. Hardware integration with an FPGA, and implementation in an electro-optical media, will result in orders of magnitude speed-ups to gigahertz speeds. If a design were optimized for usage as a PUF with the proper, previously mentioned controls, we postulate that the existence of reconfigurable optical PUFs will greatly enhance the security of future optical communications.

\section{Conclusion} %sec V
The PIC device shown in this work meets the criteria for a weak PUF given by \cite{herder14} and appears to also satisfy the definition of a strong PUF. The rapidly expanding research on large scale interferometric PICs, and the wide fields in which they are suggested for use, implies that such devices may become ubiquitous in the near future. This work shows that such large integrated devices carry with them useful amounts of unique randomness that can be used for tasks such as device identification, authentication, and other cryptographic tasks.

\section*{Acknowledgments}
The authors would like to acknowledge the group of D. Englund at MIT for assistance in the design and fabrication of the experimental optical chip. A. M. Smith would like to thank N. Stolten and F. H. Long of ARDEC for initial discussions prompting this line of study. H S. Jacinto would like to thank AFRL for doctoral fellowship support. Any opinions, findings, conclusions, or recommendations expressed in this material are those of the authors and do not necessarily reflect the views or endorsement of the U.S. Air Force Research Laboratory.

\bibliographystyle{IEEEtran}
\bibliography{puf_paper_bib}

\vskip -1.25\baselineskip plus -1fil

\begin{IEEEbiographynophoto}
{A.~Matthew~Smith} received his Ph.D. from Tulane University, New Orleans, Louisiana, USA, in 2010, where he worked on optimization of linear optics. He received a National Research Council fellowship with the Air Force Research Laboratory (AFRL) to work on cluster-state quantum computing. He completed post-doctorate research at Oak Ridge National Laboratory that focused on experimental implementations of quantum key distribution protocols. He is currently a full-time Senior Physicist at the AFRL, Information Directorate, Quantum Information Science group. His current work and research focuses on quantum frequency conversion and integrated optical devices from concept/design to fabrication and packaging.
\end{IEEEbiographynophoto}

\vskip -1.25\baselineskip plus -1fil

\begin{IEEEbiographynophoto}
{H~S.~Jacinto} (S'13 -- M'18) received his Ph.D. in electrical and computer engineering from Boise State University, Boise, Idaho, USA, in 2020, where he worked on hardware acceleration and SoC design of secure protocols. From 2015 to 2017 he worked with Idaho National Lab and the Advanced Energy Lab researching secure communications and sensor design for nuclear reactor monitoring. From 2018 to present, he has been with the Air Force Research Laboratory, Quantum Information Science group under a fellowship researching quantum information processing systems, integrated quantum photonics, and quantum control. His research focuses on quantum network cybersecurity, quantum informatics, and secure adaptive hardware anti-tamper and encryption technologies.
\end{IEEEbiographynophoto}

\end{document}